\documentclass[11pt,onecolumn,amssymb,nofootinbib]{revtex4}
\usepackage{amsmath, amsthm, amscd, amssymb}
\usepackage{bm}
\usepackage{bbm}

\begin{document}
\title{\bf Bulk Heating Effects as Tests for  Collapse Models}

\author{Stephen L. Adler}
\email{adler@ias.edu} \affiliation{Institute for Advanced Study,
Einstein Drive, Princeton, NJ 08540, USA.}
\author{Andrea Vinante}
\email{A.Vinante@soton.ac.uk}\affiliation{Department of Physics and Astronomy,
University of Southampton, SQ17 1BJ, United Kingdom}

\begin{abstract}
We discuss limits on the noise strength parameter in mass-proportional-coupled wave function collapse models implied by bulk heating effects, and examine the role of the noise power spectrum in comparing experiments of different types. This comparison
utilizes a calculation of the rate of heating through phonon excitation implied by
a general noise power spectrum $\lambda(\omega)$.   We find that in the standard heating formula,
the reduction rate $\lambda$ is replaced by
$\lambda_{\rm eff}=\frac{2}{3 \pi^{3/2}} \int d^3w e^{-\vec w^2} \vec w^2 \lambda(\omega_L(\vec w/r_c))$,
 with $\omega_L(\vec q)$ the  longitudinal acoustic phonon frequency as a function of wave number $\vec q$, and with $r_C$ the noise
correlation length.  Hence if the noise power spectrum is cut off below $\omega_L(|\vec q| \sim r_c^{-1})$,  the bulk heating rate is sharply reduced, allowing compatibility
of current experimental results.
We suggest possible
new bulk heating experiments that can be performed subject to limits placed by natural heating from radioactivity and cosmic rays. The proposed experiments exploit the vanishing of thermal transport in the low temperature limit.

\end{abstract}
\maketitle
\section{Introduction and comparison of bounds on the effective noise coupling}

There is increasing interest in testing wave function collapse models \cite{collapse}, and in particular the continuous spontaneous localization (CSL) model,  by searching for effects associated with the small noise which drives wave function collapse when nonlinearly coupled in the Schr\"odinger
equation.  The original proposals for the noise coupling strength were so small that devising
suitable experiments was problematic, but the situation has changed with the suggestion \cite{adler1}   that latent image formation, such as deposition of a developable track in an emulsion or in an etched track detector, already constitutes a measurement embodying wave function collapse and so requires an enhanced noise coupling. A recent
cantilever experiment of Vinante et al. \cite{vin}  has set bounds consistent with the  parameters suggested in \cite{adler1}, and reports a possible noise signal. Thus, it is timely
to consider further experiments  \cite{alternative} which could detect or rule out a noise coupling with
the strength suggested by \cite{vin}.

For a body comprised of a group of particles of total mass $M$, the center-of-mass energy gain for white noise with mass-proportional coupling is given by the standard formula \cite{gainref}
\begin{equation}\label{engain}
\frac{dE}{dt}=\frac{3}{4}\lambda \frac{\hbar^2}{r_C^2} \frac{M}{m_N^2}~~~,
\end{equation}
with $m_N$ the nucleon mass and $\lambda$ the coupling parameter for white noise with noise correlation length $r_C$ and no frequency spectrum cutoff.  Dividing by $M$, Eq. \eqref{engain}
can be rewritten as a formula for the energy gain rate per unit mass,
\begin{equation}\label{engain1}
\frac{dE}{dt\,dM}=\frac{3}{4}\lambda \frac{\hbar^2}{r_C^2} \frac{1}{m_N^2}~~~.
\end{equation}
For the noise coupling parameter $\lambda=10^{-7.7} {\rm s}^{-1}$ suggested in \cite{vin},  and the conventional value $r_C=10^{-5} {\rm cm}$,
Eq. \eqref{engain1} corresponds to
\begin{equation}\label{engain3}
\frac{dE}{dt\,dM}\simeq 40\frac{ {\rm MeV}}{{\rm g ~s}} \simeq 0.64  \times 10^{-8}  \frac{{\rm W}}{{\rm kg}}~~~.
\end{equation}
Since heating rates of $100 {\rm pW}/{\rm kg}= 10^{-10} {\rm W}/{\rm kg}$ are attained in low temperature experiments \cite{pobell}, with the limit accounted for by modeling
energy deposition from radioactive decays and penetrating muons \cite{nazar}, the residual heating from unknown sources is limited to roughly $10^{-11} {\rm W}/{\rm kg}$.
This means that the effective $\lambda$ for bulk heating is at most $3.1 \times 10^{-11} {\rm s}^{-1}$, ruling out a collapse model white noise interpretation of the
excess noise reported in \cite{vin}.

A similar bound on the bulk heating rate is given by Earth's energy balance.  Table 6.3 of de Pater and Lissauer \cite{depater} gives the luminosity to mass ratio for solar system objects,
with a value $6.4 \times 10^{-12} {\rm W}/{\rm kg}$ for Earth.  Estimates of primordial and radiogenic sources of Earth heat roughly account for this, but have
uncertainties that could allow   $\sim 3\times 10^{-12} {\rm W} /{\rm kg}$ to come from unknown heating sources.  This places a limit of $ \sim 10^{-11} {\rm s}^{-1}$ on the
effective $\lambda$ for bulk heating.   There is a caveat here, because
as noted in \cite{adler1}, when the effects of dissipation are included, as in the model of Bassi, Ippoliti, and Vacchini \cite{ippoliti}, the rate of heat production can vanish at large times where a limiting temperature is reached. For example, with the parameters of \cite{ippoliti},  a limiting temperature of $0.1$ K is reached on a time scale of billions of years, giving a current noise-induced Earth  heat production rate smaller than given by Eq. \eqref{engain1}, permitting a larger effective $\lambda$.

The effective CSL model $\lambda$ for bulk heating experiments can be strongly reduced if the noise is non-white, with a power spectrum cutoff.  A spectral energy cutoff is already suggested by experimental limits on spontaneous gamma ray emission from germanium \cite{germ}, which shows that the noise strength suggested in \cite{adler1} is ruled out unless
the noise power spectrum cuts off at an angular frequency below $\sim 15 \,{\rm keV}/\hbar \sim 2 \times 10^{19} {\rm s}^{-1}$.  For bulk heating of solids the noise
couples through longitudinal acoustic phonon excitation, and with $r_c\sim 10^{-5} {\rm cm}$,  heating  takes place only if the noise has frequency components of at least $\omega_L(|\vec q|=r_c^{-1} )\sim v_s |\vec q| \sim 0.4 \times 10^{11} {\rm s}^{-1}$, with $\omega_L(\vec q)$ the longitudinal phonon frequency at wave number $\vec q$, and $v_s$ the speed of sound (which for this estimate we have taken as $4000 \, {\rm m}/{\rm s}$ characteristic of copper at low temperature).  A  quantitative calculation given in
the Appendix  shows that for non-white noise with power spectrum $\lambda(\omega)$ heating a solid by phonon excitation, Eq. \eqref{engain1} is replaced by
\begin{equation}\label{engain4}
\frac{dE}{dt\,dM}=\frac{3}{4}\lambda_{\rm eff} \frac{\hbar^2}{r_C^2} \frac{1}{m_N^2}~~~,
\end{equation}
with $\lambda_{\rm eff}$ given by
\begin{equation}\label{lameff}
\lambda_{\rm eff}= \frac{2}{3 \pi^{3/2}} \int d^3w e^{-\vec w^2} \vec w^2 \lambda(\omega_L(\vec w/r_c))~~~.
\end{equation}
When $\lambda(\omega)$ is a constant independent of $\omega$, Eq. \eqref{lameff} reduces to $\lambda_{\rm eff}=\lambda$, and Eq. \eqref{engain1} is recovered, but when
there is a frequency cutoff below the phonon excitation frequency, the effective noise coupling is strongly reduced. Thus, if the noise reported in \cite{vin} at the
very low cantilever frequency of $8174 {\rm s}^{-1}$ were due to CSL,
 it would be  further evidence for non-white CSL model noise.

\section{Alternative experiments for bounding $\lambda_{\rm eff}$}
Assuming now a  maximum value $\lambda_{\rm eff}\sim  10^{-11} {\rm s}^{-1}$ consistent with low temperature experiments and non-dissipative Earth heating, let us explore possible
alternative experiments for detecting or further bounding $\lambda_{\rm eff}$.  {\it Feasibility of such experiments assumes (a) that the background heating rate from cosmic ray muons and radioactive decays can be reduced much below $3 \times 10^{-12} {\rm W}/{\rm kg}$ by shielding, underground operation, and careful choice of materials, and (b) all other known sources of heat leaks in ultralow temperature experiments, such as vibrations, relaxation from two-level systems, and hydrogen ortho-para conversion, have been suppressed.} To present estimates we multiply
$3 \times 10^{-12} {\rm W}/{\rm kg}$  by  the solid density $\rho$, so that the maximum allowed  value of $\lambda_{\rm eff}$ corresponds to a volume heating rate of
\begin{equation}\label{heating}
H=3\times 10^{-15} \rho {\rm W}/{\rm cm}^3 ~~~,
\end{equation}
with $\rho$ the density in units ${\rm g}/{\rm cm}^3$.  We consider two geometries for
which the heat transport problem is effectively one dimensional and easily solved.

\begin{enumerate}
\item  For a sphere of radius $R$ cm and density $\rho$, the total heating rate  is $4\pi R^3 H/3$, and so the steady state rate of escape of heat from unit area
of the surface will be
\begin{equation}\label{spherheat}
\dot Q_{\rm sphere}= R H/3
\end{equation}
in units ${\rm W}/{\rm cm}^2$.  This must balance the rate of transport of heat per unit area from the surface of the sphere, at temperature $T_1$, to the surrounding cryostat surfaces, at temperature $T_2$.  This is given \cite{pobell} in units ${\rm W}/{\rm cm}^2$ by the formula
\begin{equation}\label{transport}
\dot{Q}_{\rm transport} = 5.67 \times 10^{-12}\epsilon  (T_1^4-T_2^4)[K]^4+ 0.02 a P[{\rm mbar}] (T_1-T_2)[K]~~~,
\end{equation}
with the first term the Stefan-Boltzmann equation for radiative heat transfer, and the second term, which is linear in the pressure $P$,  coming from gas particle conduction in the cryostat.  Here  $a \leq 1$ is an ``accommodation coefficient'' for gas particles on the cryostat walls, which can be as small as 0.02 for a clean metal surface in contact with helium gas, and $\epsilon \leq 1$ is the emissivity for radiative transfer.   The steady state  surface temperature of the sphere $T_1$ is determined by equating $\dot{Q}_{\rm sphere}$ to $\dot{Q}_{\rm transport}$.
Taking as an example
the density of lead $\rho = 11.4 \,{\rm g}/{\rm cm}^3$, and a sphere of radius $R=50\, {\rm cm}$ (which would fit in the CUORE underground experiment cryostat \cite{cuore}) , we have
\begin{equation}\label{sphere1}
\dot{Q}_{\rm sphere}=5.7 \times 10^{-13}~~~,
\end{equation}
while taking $a=0.02$ and $P=10^{-6}{\rm mbar}$ gives
\begin{equation}\label{transp}
\dot{Q}_{\rm transport} = 5.67 \times 10^{-24} \epsilon (T_1^4-T_2^4)[{\rm mK}]^4+4 \times 10^{-13} (T_1-T_2) [{\rm mK}]~~~,
\end{equation}
both in units $ {\rm W}/{\rm cm}^2$. Evidently, for millikelvin $T_1$ and $T_2$  radiative heat transfer is completely negligible and Eqs. \eqref{sphere1} and \eqref{transp} are of similar size at the relatively high pressure of $10^{-6}$ mbar.
 In real millikelvin cryostats the actual residual pressure is typically orders of magnitude lower than $10^{-6}$ mbar because the vapor pressure of anything, including helium, drops to zero exponentially with decreasing temperature.
 The  heating rate H can therefore be estimated by measuring the steady state surface temperature as a function of the gas pressure.  Fast and sensitive measurements of temperature in the range of a few mK can be readily done using resistive probes. For high accuracy, SQUID-based noise thermometry has been recently demonstrated to be very effective down to 42 $\mu$K \cite{roth}.

Similar reasoning gives the temperature distribution inside a sphere of material with thermal conductivity $k(T)$.  At a given distance $r$ from the center of the
sphere, the energy transport rate through the spherical surface of radius r is equal to
\begin{equation}\label{spherout}
E_{\rm out}= -4 \pi r^2 k(T) \frac{dT}{dr}~~~,
\end{equation}
which in steady state  must balance the heating rate of the volume within radius $r$,
\begin{equation}\label{spherin}
E_{\rm in}=\frac{4 \pi}{3} r^3 H~~~,
\end{equation}
giving the differential equation
\begin{equation}\label{diffeq}
-k(T)\frac{dT}{dr}=\frac{1}{3} r H~~~.
\end{equation}
Integrating from the center of the sphere at radius 0  to radius $R$, with respective temperatures $T_c$ and $T_1$, this gives
\begin{equation}\label{int}
-\int_{T_c}^{T_1} \, k(u)du= \frac{R^2 H}{6}~~~.
\end{equation}
For $k(u)=\hat k_0 u^{\beta}$,
this becomes
\begin{equation}\label{int1}
-\frac{\hat k_0}{1+\beta} (T_1^{1+\beta}-T_c^{1+\beta})=\frac{R^2 H}{6}~~~,
\end{equation}
 which gives
\begin{equation}\label{intt1}
 (T_c^{1+\beta}-T_1^{1+\beta})^{1/(1+\beta)}=\left[\frac{1+\beta}{\hat k_0K^{1+\beta}} \frac{R^2 H}{6}\right]^{1/(1+\beta)}[{\rm K}]~~~.
\end{equation}
 To give a numerical estimate, for the good thermal insulator Torlon 4203 \cite{torlon}, with density $1.42~ {\rm g~cm^{-3}}$ and with $k(T)=6.13 \times 10^{-3} (T/{\rm K})^{2.18} {\rm W/(m~K)}$, so that $\beta=2.18$ and $\hat k_0 {\rm K}^{3.18}= 6.13 \times 10^{-3} {\rm W}/{\rm m}$, the right hand side of  Eq. \eqref{intt1} for $R=50 \,{\rm cm}$  is 6.1 mK.
 \medskip

We have used Torlon as a convenient example for estimates, but probably it would not be the most suitable material for these  experiments. As discussed by Pobell \cite{pobell}, amorphous materials are usually rich in two-level systems leading to slow relaxation processes with time-dependent heat release. Moreover, polymers and plastics like Torlon may easily absorb impurities which can also give unreliable thermal properties.  For realistic
 experiments, it would be better to use crystalline insulators (such as sapphire or silicon) or superconducting metals. In both cases the thermal conductivity is much higher than that of Torlon at $T>100~ {\rm mK}$, but drops as $T^3$ and approaches  the conductivity of plastic materials like Torlon in the mK range.

Concerning the feasibility of this experiment, we note that spheres of meter size have already been suspended and cooled down to millikelvin temperature. An example is Minigrail, a CuAl spherical gravitational wave detector, which  was suspended through a central rod suspension and cooled in vacuum  down to 60 mK within a few days time \cite{waard}. In that case, the dominant residual heating mechanism was likely relaxation of defects and hydrogen ortho-para conversion. Both issues can be avoided in a dedicated experiment by a suitable choice of materials.

\item  Another geometry with easily solvable heat transfer  would use a long cylinder (or parallelepiped, or more generally a rod of uniform cross section) of length $L$  with the ``near'' end fastened to a heat sink that provides  a large enough heat transport rate so
that the heat transport from all other surfaces given by Eq. \eqref{transport} can be ignored.  Then the heat transport problem is one dimensional, and the analog of Eq.
\eqref{diffeq} is the differential equation
\begin{equation}\label{zdiffeq}
k(T)\frac{dT}{dz}=(L-z)H~~~.
\end{equation}
Again taking $k(u)=\hat{k}_0u^{\beta}$ and integrating,  the analog of
Eq. \eqref{intt1} relating the ``far'' to the ``near'' end temperatures at $z=L$ and
$z=0$ respectively is
\begin{equation}\label{intt2}
 (T_{\rm far}^{1+\beta}-T_{\rm near}^{1+\beta})^{1/(1+\beta)}=\left[\frac{1+\beta}{\hat k_0K^{1+\beta}} \frac{L^2 H}{2}\right]^{1/(1+\beta)}[{\rm K}]~~~.
\end{equation}
For a rod  with the parameters quoted above, and $L=50\,{\rm cm}$, the right hand side of Eq. \eqref{intt2} is 8.6 mK.

In a  variant of the rod geometry, one can attach to
the ``far'' end of the rod an object of larger size,  which acts as a ``CSL noise absorber''. The CSL heat released in the absorber can be much larger than that in the rod.   The temperature at the ``far'' end of the rod will then be determined by
matching the heat flow per unit area within the rod with the heat flow per unit area $\dot{Q}_{\rm ABS}$ entering the rod from the absorber, so that Eq. \eqref{zdiffeq} becomes
\begin{equation}\label{zdiffeq1}
k(T)\frac{dT}{dz}=(L-z)H+\dot{Q}_{\rm ABS}~~~,
\end{equation}
which on integration gives
\begin{equation}\label{intt3}
 (T_{\rm far}^{1+\beta}-T_{\rm near}^{1+\beta})^{1/(1+\beta)}=\left[\frac{1+\beta}{\hat k_0K^{1+\beta}}\left( \frac{L^2 H}{2}+\dot{Q}_{\rm ABS}L \right)\right]^{1/(1+\beta)}[{\rm K}]~~~.
\end{equation}
  This provides the freedom of independently tuning the rod thermal conductivity (by making the cross section arbitrarily small) and the heat input from CSL (by choosing the  material and size of the CSL absorber). From an experimental point of view this gives a very flexible design. Of course, one must make sure that the heat transport mechanisms of Eq. \eqref{transport} from the surfaces of both the rod and the absorber are kept negligible, as discussed above following Eq. \eqref{transp}.

\end{enumerate}

To conclude, current bounds on the effective noise coupling $\lambda_{\rm eff}$ for bulk heating of solids show that in designing experiments to test the enhanced rate \cite{adler1}, \cite{vin} that makes latent image formation a measurement, it will be important to take the power spectrum of the noise into account.  Because thermal transport rates vanish at zero temperature, millikelvin and submillikelvin experiments to further improve the bounds on $\lambda_{\rm eff}$ by one or two orders of magnitude may be feasible.  However, underground operation is probably necessary in order to evade the limiting heating rate from cosmic rays.
\section{Acknowledgements}
We wish to thank Angelo Bassi for helpful comments.  A.V. acknowledges support from
EU FET project TEQ (grant agreement 766900). The authors thank the referee for a careful reading of the paper and helpful comments.

The appendix (except for the addition of the paragraph containing
Eq. \eqref{lambdaest}) is based on the
preprint \cite{adler2} by one of the authors (SLA).  The other author
(AV) later noted a preprint by M. Bahrami \cite{bahrami}  in which the calculation is done by a non-perturbative
method.

\appendix
\section{Heating through phonon excitation implied by collapse models}

In this appendix we calculate the heating rate of a solid through phonons excited by
CSL noise with a non-white power spectrum.
\subsection{Monatomic lattice unit cell}
Consider a system in initial state $i$ with energy $E_i= \hbar \omega_i$ at time $t=0$, acted on by a perturbation $V$ which at time $t$ leads to a transition to
a state $f$ with energy $E_f=\hbar \omega_f$.  Working in the interaction picture, the transition amplitude $c_{fi}(t)$ is given by
\begin{equation}\label{tranamp}
\langle f|c(t)|i \rangle \equiv c_{fi}(t)= -\frac{i}{\hbar}\int_0^t V_{fi}(t^{\prime})e^{i\omega_{fi}t^{\prime}}dt^{\prime}~~~,
\end{equation}
with $\omega_{fi}=\omega_f-\omega_i$.  For $V$ we take the noise coupling in the mass-proportional continuous spontaneous localization (CSL) model,
\begin{align}\label{noise}
V=&\int d^3z \frac{dW_t(\vec z)}{dt} {\cal V}(\vec z,\{\vec x\}) ~~~,\cr
{\cal V}(\vec z,\{\vec x\})=&-\frac{\hbar}{m_N} \sum_{\ell} m_{\ell} g(\vec z-\vec x_{\ell} )~~~,
\end{align}
where we have followed the notation used in \cite{adlerram}.  Here $\vec x_{\ell} $ are the coordinates of atoms of mass $m_{\ell}$, $g(\vec x)$ is a spatial correlation function, conventionally taken as a Gaussian
\begin{equation}\label{correl}
g(\vec x)=(2\pi)^{-3/2}\, (r_c)^{-3} e^{-{\vec x}^2/(2 r_c^2)}=(2\pi)^{-3} \int d^3q e^{-r_c^2 \vec q^{\,2}/2\,-i\vec q \cdot \vec x}~~~,
\end{equation}
and the non-white noise has expectation ${\cal E}$
\begin{equation}\label{noiseex}
{\cal E}\left[\frac{dW_t(\vec x)}{dt}\, \frac{dW_{t^{\prime}}(\vec y)} {dt^{\prime}}\right] =\frac{1}{2\pi} \int_{-\infty}^{\infty} d\omega
\gamma(\omega) e^{-i\omega(t-t^{\prime})}\delta^3(\vec x-\vec y)~~~,
\end{equation}
with $\gamma(\omega)=\gamma(-\omega)$ related to the reduction rate parameter $\lambda(\omega)$ by
\begin{equation}\label{gamal}
\gamma(\omega)=8\pi^{3/2} r_c^3\lambda(\omega)~~~.
\end{equation}

We wish now to calculate the expectation ${\cal E}[E(t)]$ of the energy attained by the system at time $t$, given by
\begin{equation}\label{enex}
{\cal E}[E(t)]={\cal E}[\sum_f \hbar \omega_{fi}|c_{fi}(t)|^2]~~~.
\end{equation}
Substituting Eqs. \eqref{tranamp} -- \eqref{gamal}, carrying out integrations, and using the formulas \cite{cohen}
\begin{align}\label{deltaform}
\int_0^t dt^{\prime} e^{i(\omega_{fi}-\omega) t^{\prime}} =& \frac{e^{i(\omega_{fi}-\omega) t}-1}{i (\omega_{fi}-\omega)}
\equiv 2\pi  e^{i(\omega_{fi}-\omega) t/2}\delta^{(t)}(\omega_{fi}-\omega)~~~,\cr
[\delta^{(t)}(\omega_{fi}-\omega)]^2\simeq &\frac{t}{2\pi} \delta^{(t)}(\omega_{fi}-\omega)~~~,\cr
\end{align}
we find in the large $t$ limit the formula for the energy gain rate
\begin{equation}\label{rate}
t^{-1} {\cal E}[E(t)]=\frac{r_c^3}{\pi^{3/2}m_N^2} \int d^3 q \sum_f e^{-r_c^2 \vec q^2} \lambda(\omega_{fi})\hbar\omega_{fi}
\big|\langle f|\sum_{\ell} m_{\ell} e^{i\vec q \cdot \vec x_{\ell} }|i\rangle \big|^2~~~.
\end{equation}

The next step is to evaluate the matrix element appearing in Eq. \eqref{rate} by introducing  phonon physics, following the exposition in the text of Callaway \cite{cal}.  We
consider first the simplest case of a monatomic lattice with all $m_{\ell} $ equal to $m_A$, independent of the index $\ell$, and write the atom coordinate $\vec x_{\ell}$ as
\begin{equation}\label{coord}
\vec x_{\ell}=\vec R_{\ell}+\vec u_{\ell}~~~,
\end{equation}
with $\vec R_{\ell}$ the equilibrium lattice coordinate and with $\vec u_{\ell}$ the lattice displacement induced by the noise perturbation.   Writing
\begin{equation}\label{internsum}
\sum_{\ell} m_{\ell} e^{i\vec q \cdot \vec x_{\ell} }=m_A \sum_{\ell} e^{i\vec q \cdot  \vec R_{\ell}}e^{i\vec q \cdot  \vec u_{\ell} }~~~,
\end{equation}
we note that since the Gaussian in Eq. \eqref{rate} restricts the magnitude of $\vec q$ to be less than of order of $r_c^{-1}$, with $r_c \sim 10^{-5} {\rm cm}$, whereas
the magnitude of the lattice displacement is much smaller than $10^{-8} {\rm cm}$, the exponent in $e^{i\vec q \cdot  \vec u_{\ell} }$ is a very small quantity.
So we can Taylor expand to write
\begin{equation}\label{expan}
e^{i\vec q \cdot  \vec u_{\ell} }\simeq 1+ i\vec q \cdot  \vec u_{\ell}~~~.
\end{equation}
The leading term $1$ does not contribute to energy-changing transitions, so we have reduced the matrix element in Eq. \eqref{rate} to the simpler form
\begin{equation}\label{matrix1}
\langle f|\sum_{\ell} m_{\ell} e^{i\vec q \cdot \vec x_{\ell} }|i\rangle\simeq im_A \langle f| \sum_{\ell} e^{i\vec q \cdot  \vec R_{\ell}} \vec q \cdot \vec u_{\ell} |i\rangle
~~~,~~~~f \neq i~~~.
\end{equation}
The approximation leading to Eq. \eqref{matrix1}  is a phonon analog of the electric dipole approximation made in electromagnetic radiation rate calculations.

We now substitute the expression \cite{cal} for the lattice displacement in terms of phonon creation and annihilation operators,
\begin{equation}\label{disp}
\vec u_{\ell}= \frac{\Omega}{8\pi^3} \left(\frac{\hbar {\cal N}}{m_A}\right)^{1/2} \sum_j \int \frac{d^3 k}{(2\omega_j(\vec k))^{1/2}}
\big[\vec e^{\,(j)}(\vec k)   e^{i\vec k \cdot \vec R_{\ell}} a_j(\vec k)+ \vec e^{\,(j)*}(\vec k) e^{-i\vec k \cdot \vec R_{\ell}} a_j^{\dagger}(\vec k)\big]~~~,
\end{equation}
where the sum on $j$ runs over the acoustic phonon polarization states, and where $\Omega$ and ${\cal N}$ are respectively the lattice unit cell volume, and
 the number of unit cells.  Taking the initial state $i$ to be the zero phonon state, only the $a_j^{\dagger}$ term
in Eq. \eqref{disp} contributes, and we can evaluate the sum over lattice sites $\ell$ in Eq. \eqref{matrix1} using the formula \cite{cal}
\begin{equation}\label{sum1}
\sum_{\ell} e^{i (\vec q-\vec k)\cdot \vec R_{\ell}}= \frac{8\pi^3}{\Omega}\delta^3(\vec q-\vec k)~~~.
\end{equation}
Carrying out the $\vec k$ integration, noting that $\vec q \cdot \vec e^{\,(j)}(\vec q)$ selects the longitudinal phonon with frequency $\omega_L(\vec q)$, defining $\vec w=r_c \vec q$, writing $M={\cal N} m_A$ for the total system mass, and assembling all the pieces, we arrive at the answer
\begin{align}\label{final1}
t^{-1} {\cal E}[E(t)]=&\frac{\hbar^2 M}{m_N^2 r_c^2} \frac{1}{2 \pi^{3/2}} \int d^3w e^{-\vec w^2} \vec w^2 \lambda(\omega_L(\vec w/r_c))=\frac{3}{4} \frac{\hbar^2\lambda_{\rm eff}  M}{m_N^2 r_c^2}~~~,\cr
\lambda_{\rm eff}\equiv &\frac{2}{3\pi^{3/2}} \int d^3w e^{-\vec w^2} \vec w^2 \lambda(\omega_L(\vec w/r_c))~~~.\cr
\end{align}
In the white noise case, where  $\lambda(\omega)$ is a constant $\lambda$, we can pull it outside the $\vec w$ integral and use
\begin{equation}\label{pulled}
\int d^3 w e^{-\vec w^2} \vec w^2 =\frac{3}{2} \pi^{3/2}
\end{equation}
to get the standard formula \cite{gainref}
\begin{equation}\label{stdform}
t^{-1} {\cal E}[E(t)] = \frac{3}{4} \frac{\hbar^2 \lambda M}{m_N^2 r_c^2}~~~.
\end{equation}

When the noise spectrum has a cutoff below $\omega_L(\vec q)$ for $|\vec q| \sim r_c^{-1}$, the energy gain rate is sharply reduced. To estimate this, let us assume a Gaussian frequency cutoff of the form $\lambda(\omega)=\lambda \exp(-\omega^2 t_c^2)$, with $t_c$ a correlation time, and take $\omega_L(\vec q)=
v_s |\vec q|$, with $v_s$ the sound velocity in the solid.  Then Eq. \eqref{final1} gives
\begin{equation}\label{lambdaest}
\lambda_{\rm eff}=\frac{\lambda}{(1+v_s^2 t_c^2/r_c^2)^{5/2}}~~~.
\end{equation}
Thus, $\lambda \simeq 10^{-7.7} {\rm s}^{-1}$ and $\lambda_{\rm eff} \leq 10^{-11}  {\rm s}^{-1}$  would correspond to $r_c/t_c \leq v_s/4.5$, strongly ruling out a naive guess
$r_c/t_c \simeq {\rm light~velocity}$.  When Eq. \eqref{lambdaest} is approximated as
$\lambda_{\rm eff} \simeq \lambda (r_c/(v_s t_c))^5$, it corresponds (up to constant factors)  to the estimate given following Eq. (45) of Bahrami \cite{bahrami} which assumes  a step function frequency cutoff.

\subsection{Extensions to multi-atom unit cell and nonzero phonon initial state}

Although we have derived the result of Eq. \eqref{final1} for the case of a monatomic lattice and a zero phonon initial state, the result is more general as we shall now
show. In the monatomic case, focusing  only on the atomic mass factors and longitudinal phonon polarization vectors, Eqs. \eqref{matrix1} and \eqref{disp} give a factor
\begin{equation}\label{mono}
m_A^{1/2} \vec{e}^{\,(L)*}(\vec k)\simeq m_A^{1/2} \vec{e}^{\,(L)*}(\vec 0)~~~.
\end{equation}
After the $\simeq$ sign we have used the fact, noted after Eq. \eqref{internsum},
that the correlation length $r_C$ allows only contributions from phonon wavelengths that are long on a lattice scale, corresponding to $\vec k \simeq \vec 0$.  In the multi-atom case, focusing only on acoustic phonons,\footnote{Optical phonons leave the unit cell
center of mass stationary, so obey $\sum_{\kappa}m_{\kappa}^{1/2} \vec{e}_{\kappa}^{\,(s)}(\vec 0) =0$ for any optical phonon mode $s$. Hence for mass-proportional
noise coupling, optical phonons do not contribute to the energy gain rate to leading
order in $a/r_C$, with $a$ the unit cell dimension.}  the left-hand side of Eq. \eqref{mono} is replaced by
\begin{equation}\label{multi}
m_{\kappa}^{1/2} \vec{e}_{\kappa}^{\,(L)*}(\vec k)~~~,
\end{equation}
corresponding to Eqs. (1.4.22a,b) of  \cite{cal}, with $\kappa$ labeling an atom in the
multi-atom unit cell.  Referring now to the unnumbered equation in Callaway \cite{cal} between his  Eqs. (1.1.22) and (1.1.23), which we write \big(using the fact that for $\vec k=0$ the polarization vectors are real numbers; see Callaway Eq. (1.1.21)\big) as
\begin{equation}\label{polvec}
m_{\kappa}^{-1/2} \vec{e}_{\kappa}^{\,(L)*}(\vec 0)=m_{\kappa}^{-1/2} \vec{e}_{\kappa}^{\,(L)}(\vec 0)=\vec C~~~,
\end{equation}
with $\vec C$ a constant,
we see that the longitudinal polarization vectors are no longer unit normalized, as
in the monatomic case.  Instead, the normalization is given in Eq. (1.1.18a) of
\cite{cal},
\begin{equation}
\sum_{\kappa} \vec{e}_{\kappa}^{\,(L)*}(\vec 0) \cdot \vec{e}_{\kappa}^{\,(L)}(\vec 0)
=1~~~,
\end{equation}
which on substituting Eq. \eqref{polvec} gives
\begin{equation}\label{cmag}
|\vec C|= \left(\sum_{\kappa} m_{\kappa}\right)^{-1/2}~~~,
\end{equation}
and implies for small $\vec k$
\begin{equation}\label{polvec1}
m_{\kappa}^{1/2}\hat k \cdot  \vec{e}_{\kappa}^{\,(L)*}(\vec k) \simeq  m_{\kappa}  |\vec C| =  m_{\kappa} \left(\sum_{\kappa} m_{\kappa}\right)^{-1/2}~~~.
\end{equation}
Recalling Eqs. \eqref{expan}--\eqref{sum1}, summing over $\kappa$ to get the total contribution to the one-phonon creation amplitude, we have
\begin{equation}\label{summing}
 \sum_{\kappa} m_{\kappa} \left(\sum_{\kappa} m_{\kappa}\right)^{-1/2}~~~,
\end{equation}
which when squared gives a factor
\begin{equation}\label{squared}
\sum_{\kappa} m_{\kappa}=m_{\rm cell}~~~,
\end{equation}
which is the total atomic mass in the unit cell.   Thus the only change from
the monatomic to the multi-atomic case is the replacement of $m_A$ by $m_{\rm cell}$,
and since ${\cal N}m_{\rm cell}=M$, the total system mass, the monatomic formula
of Eq. \eqref{final1} is unchanged.  Heuristically, the reason for this is that, as
emphasized by Callaway, for $\vec k=0$ acoustic phonons Eq. \eqref{polvec} implies that
all ``...particles in each unit cell move in parallel with equal amplitudes'', and so
behave as a single particle with mass $m_{\rm cell}$.

The above derivation assumes an initial state with no phonons, but this assumption
is not needed to get Eq. \eqref{final1}.
When the initial state is constructed from  $n$-phonon states,
as in a thermal ground state, the $a^{\dagger}$ term in Eq. \eqref{disp} contributes a term proportional to $(n+1)\omega_L$ to the
energy gain, while the $a$ term in Eq. \eqref{disp} contributes a corresponding term proportional to $-n \omega_L$ to the energy gain; the sum of the two terms
is proportional to $(n+1-n) \omega_L = \omega_L$, so $n$ drops out and the formula of Eq. \eqref{final1} is recovered.  This simplification  could have been anticipated
from our earlier analysis of the noise-induced energy gain by an oscillator \cite{adlerosc}, which showed that the rate of energy gain is a constant independent of the number of
oscillator quanta that are present.

\end{document}